\documentclass[a4paper]{article}

 \usepackage{graphicx}

\usepackage[tight]{subfigure}

\usepackage{amssymb}

\makeatletter
\renewcommand\thefigure{\@arabic\c@figure}

\long\def\@makecaption#1#2{%
\vskip\abovecaptionskip
\sbox\@tempboxa{#1.\hskip 1em#2}%
\ifdim \wd\@tempboxa >\hsize
  #1.\hskip 1em#2\par
\else
  \global \@minipagefalse
  \hb@xt@\hsize{\hfil\box\@tempboxa\hfil}%
\fi \vskip\belowcaptionskip} \makeatother

\begin{document}

\title{Evolutionary Prisoner's Dilemma Game in Flocks}
\author{Zhuo Chen$^{1*}$, Jianxi Gao$^1$, Yunze Cai$^2$ and Xiaoming Xu$^{1,2,3}$\\
$^1$Shanghai Jiao Tong University, Shanghai, China\\
$^2$University of Shanghai For Science and Technology, Shanghai, China\\
$^3$Shanghai Academy of Systems Science, Shanghai, China\\
$^*$jeffchen\_ch@yahoo.com.cn}
\maketitle

\begin{abstract}

We investigate an evolutionary prisoner's dilemma game among
self-driven agents, where  collective motion of biological flocks is
imitated through averaging directions of neighbors. Depending on the
temptation to defect and the velocity at which agents move, we find
that cooperation can not only be maintained in such a system but
there exists an optimal size of interaction neighborhood, which can
induce the maximum cooperation level. When compared with the case
that all agents do not move, cooperation can even be enhanced by the
mobility of individuals, provided that the velocity and the size of
neighborhood are not too large. Besides, we find that the system
exhibits aggregation behavior, and cooperators may coexist with
defectors at equilibrium.
\end{abstract}

\section{Introduction}
Cooperation is commonly observed in genomes, cells, multi-cellular
organisms, social insects, and human society, but Darwin's theory of
evolution implies fierce competition for existence among selfish and
unrelated individuals.
In past decades, much effort has been devoted to understanding the
mechanisms behind the emergence and maintenance of cooperation.
In this context, the prisoner's dilemma game is a widely used model to
illustrate the conflict between selfish and cooperative behavior.

The traditional prisoner's dilemma (PD) game is a two-player game,
where each player can choose either cooperation (C) or defection
(D). Mutual cooperation pays each a reward $R$, while mutual
defection brings each a punishment $P$. If one player chooses to
cooperate while the other prefers to defect, the cooperator obtains
the sucker's payoff $S$ and the defector gains the temptation $T$.
The four payoff values satisfy the following conditions: $T>R>P>S$
and $2R>S+T$. According to the inequalities, defection is the
optimal strategy to maximize payoff for a selfish player in a
one-shot game, no matter what the opponent dose. But the total
income of two defectors is lower than that of two cooperators. Hence
the dilemma arises, and defection is evolutionarily stable in a
well-mixed population \cite{Hauert2004}.

The spatial PD games have attracted much attention since Nowak and
May reported the stable coexistence of cooperators and defectors in
a two-dimensional lattice \cite{Nowak1993}. After that, many works
have been done to add randomness to the deterministic game dynamics.
For example, the noise can be introduced based on the payoff
difference, which allows an inferior strategy to be followed with
certain probability \cite{Szabo1998}. The mapping of game payoffs to
individual fitness can follow different distributions, which
accounts for social diversity \cite{Perc2008}. And in the dynamic
preferential selection model, the more frequently a neighbor's
strategy is adopted by the focal player, the larger probability will
be chosen to refer to in the subsequent rounds \cite{Wu2006}.
Besides, networks describing connections among individuals have also
been also extended from lattices to complex networks
 \cite{Santos2005,Gomez-Gardenes2007,Ren2007,Chen2008,Du2009}.
For more details about spatial evolutionary games, please see Ref. \cite{Szabo2007,Nowak2006,Doebeli2005} and references therein.

In spatial games mentioned above, players are located on the
vertices of the network, and edges among vertices determine who
plays with whom. Often the network is assumed to be static. However,
in real social systems, the network size may continuously change as
individuals join or quit, and the network structure can also evolve
as links are created or broken.
It has been reported that co-evolution constitutes a key mechanism
for the sustainability of cooperation in dynamic networks \cite{Zimmermann2004,Santos2006,Szolnoki2008a,Szolnoki2008,Szolnoki2009,Fu2008}.

For a network, the movement of individuals may either change its
size or its structure. For example, when people drive, cell phones
connect with different base stations in the mobile communication
network. And moving house brings one new neighbors in the
acquaintance network. In fact, the motion of individuals is an
important characteristic of the social network \cite{Gonzalez2006},
and the patterns of human mobility have drawn much attention in the
past years \cite{Brockmann2006,Gonzalez2008}. When the spatial
structure has been introduced, it is natural to consider the
evolution of cooperation in mobile individuals.

By intuition the introduction of mobility would lead to the
dominance of defection because mobile defectors can expect more
cooperators to employ than that of the static network, and escape
retaliation of former partners by running away. Yet, the correlation
between cooperation and mobility is more complex than intuition.
Mobility could affect the origin of altruism, while the rise of
altruism cost would lead to an evolutionary reduction of mobility
\cite{LeGalliard2005}. With a win-stay, lose-shift rule cooperation
would be evolutionary stable  under generalized reciprocity
\cite{Hamilton2005}. Further, in agent-based models, mobility of
individuals can be involved explicitly as the movement of agents.
 ``Walk Away'', a simple strategy of contingent movement, can outperform complex strategies
 under a number of conditions  \cite{Aktipis2004}.
And success-driven migration may promote the spontaneous outbreak of
cooperation in a noisy world, which is dominated by selfishness and
defection \cite{Helbing2009}. Even in a blind pattern of mobility,
cooperation is not only possible but may also be enhanced for a
broad range of parameters, when compared with the case that all
agents never move \cite{Vainstein2007,Dai2007, Meloni2009}.

In the present work we study the evolution of cooperation among
mobile players, which are allowed to move in a two-dimensional plane
without periodic boundary conditions. The movement of every agent is
non-contingent, imitating the direction alignment process in
biological flocks. We find that there exists an optimal size of
interaction neighborhood, which can induce the maximum cooperation
level. When compared with the case that all agents do not move,
cooperation can not only be maintained but even be enhanced by the
movement of players. We also investigate the dependence of the
cooperator frequency on the density of agents, and the coexistence
of different strategies is illustrated.

\section{The Model}
Let $x_i(t)$ and $\theta_i(t)$, $i=1,2,...N$, denote the position
and moving direction of the agent $i$ at time $t$, $t=0,1,2...$,
respectively. Assume that each agent has the same absolute velocity
$v$. When $t=0$, all agents are randomly distributed in an $L\times
L$ square without boundary restrictions, and their directions,
$\theta_i(0)$, are uniformly distributed in $[0,2\pi)$. The position
of each agent is updated according to
\begin{equation}
  x_i(t+1)=x_i(t)+\overrightarrow{V_i}(t)\Delta t,
\end{equation}
where \overrightarrow{V_i}(t) is characterized by $v$ and
$\theta_i(t)$. In addition, $\Delta t$ is set to $1$ between two
updates on the positions.

In biological systems, such as flocks of birds and schools of fish,
individuals tend to align their moving directions with that of
nearby neighbors. To simulate the process of direction alignment in
flocks, the angle $\theta_i(t)$ of agent $i$ is updated according to
the average direction of its neighbors. Then
\begin{equation}
\theta_i(t+1)=arctan\frac{sin\theta_i(t)+\sum_{j\in
W_i(t)}sin\theta_j(t)}{cos\theta_i(t)+\sum_{j\in
W_i(t)}cos\theta_j(t)},
\end{equation}
where $W_i(t)$ denotes the neighbors set of the agent $i$ at time
$t$.

In real populations, people are believed to interact much more with
their neighbors than with those who are far away \cite{Szabo1998}.
Based on this point, when players are located on the nodes of a
fixed network, interactions often take place among immediate
players. When players are kept moving,  distances can be used to
find neighbors close to the focal one \cite{Dossetti2009}. Note in
the Vicsek model \cite{Vicsek1995},  the neighbors set $W_i(t)$ is
defined as agents within the circle of radius $r$ centered at the
agent $i$. To exclude the effects from fluctuations of the
neighborhood size, we assume that each agent will only interact
 with $k$ nearest neighbors at time $t$.
Thus $W_i(t)$ can be written as
\begin{equation} \label{neidef}
  W_i(t)=argmin_k\{{\| x_i(t)-x_j(t) \|},j\in N,j\neq i\},
\end{equation}
where the function $argmin_k\{\bullet\}$ means to find $k$ smallest
elements given in $\{\bullet\}$, and ${\|\bullet\|}$ denotes the Euclidean distance between $j$ and $i$
in the two-dimensional space.
In simulations, distances between the focal agent and the others are calculated at first. Then they are sorted in an ascending order, which means $x_1 \leq x_2 \leq x_3 \leq ... \leq x_{N-1}$. Here $x$ denote the distance between $i$ and $j$, and the suffix represents  its order.
If $x_1 \neq x_2 \neq x_3 \neq ... \neq x_k \neq x_{k+1}$, $k$ nearest agents are chosen as neighbors.
If $x_m=x_{m+1}=...=x_{m+n}$ and $1\leq m \leq k$, $k-m+1$ agents are randomly selected among $n+1$ agents when $(k-m)<n$.
The sorting process leads a directed interaction network, however.
It means $i\in W_j(t)$ does not imply $j\in W_i(t)$.
Here, $W_i(t)$ and $W_j(t)$ denote the neighbors sets of $i$ and $j$ at time $t$ respectively.

Next we introduce the evolutionary rules of our game. Initially all
players are randomly assigned one strategy of the PD with equal
probability. The strategy $s_i$ of each player can be denoted by an
unit vector $(1,0)^T$ or $(0,1)^T$, which indicates cooperation or
defection respectively. At each time step, each player plays the PD
game with his neighbors $M_i(t)$, accumulating a payoff
$P_i=\sum_{j\in M_i(t)}s_i^TAs_j$. Here we assume that
$M_i(t)=W_i(t)$. And following the common practices
\cite{Nowak1993}, the payoff matrix of the PD takes a rescaled form
as
\begin{equation}
  \mathbf{A}=\left(
  \begin{array}{ccc}
    1 & 0 \\
    b & 0
  \end{array}
  \right),
\end{equation}
where $1<b<2$. Then, every player chooses the strategy
that gains the highest payoff among itself and its neighbors at the next time step \cite{Nowak1993}.
Though the evolution of strategies and the movement of agents are characterized by two time scales respectively,
they are treated as the same here. It means that every agent modifies its position and
direction after strategy update. This process is repeated until the system reaches
equilibrium.

In our model, distances among each agent determine the network of
contacts, and the agents continuously change their positions. As a
result, the neighbors may be different at each step, though the size
of neighborhood is fixed. To characterize the evolution of the
interaction network, we calculate the new neighbors that all agents
meet at time $t$ as
\begin{equation}
  n(t)=\sum_{i=1}^{N}|W_i(t)-W_i(t)\bigcap W_i(t-1)|,
\end{equation}
where $|\bullet |$  represents the set size.

\begin{figure}[htbp]
\centering
\includegraphics[height=5cm,width=7cm]{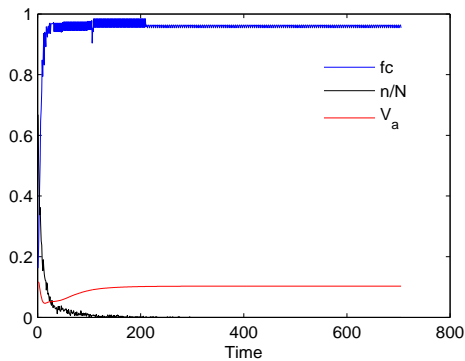}
\caption{Representative time evolutions of $fc$, $n$ and $V_a$ for $b=1.2$, $k=15$ and $v=0.05$.} \label{typicalevo}
\end{figure}

Fig. \ref{typicalevo} shows typical evolutions of $n(t)$, which is divided by $N$ for normalization,
and the frequency of cooperators $fc$.
One can find that $n(t)$ decreases to $0$ when $t>200$.
For comparison, we also plot the evolution of average normalized velocity $V_a$ defined in Ref. \cite{Vicsek1995}.
As the decrease of $n(t)$, $V_a$ also reaches a steady value,
which indicates a stable distribution of moving directions of the agents.
These findings imply that given a sufficient relaxation time, each agent owns a fixed neighborhood.
Later, we will show that without periodic boundary conditions,
the system forms many disconnected components after a long run time.
Thus the variation of neighbors, if any, would be constrained within a fraction of agents in the population.

The simulations were carried out in a system with $N=500$, $L=7$.
In each realization, we first check whether the interaction network is fixed after a suitable relaxation time.
The relaxation time is varied from $5000$ to $10^5$ generations, and the longer run time corresponds to small neighborhood size $k$ or velocity $v$.
If $n(t)\leq 1$, and this condition can hold for $q=1000$ time steps,
the network would be treated as a static one.
Then we evaluate the frequency of cooperators at equilibrium by averaging over the last 1000 generations.
All data points shown in each figure are acquired by averaging over
400 realizations of independent initial states.
\section{Results and Discussions}
\begin{figure}[htbp]
\centering \subfigure[v=0.05]{ \label{fc-b:subfig:a}
\includegraphics[height=5cm,width=5cm]{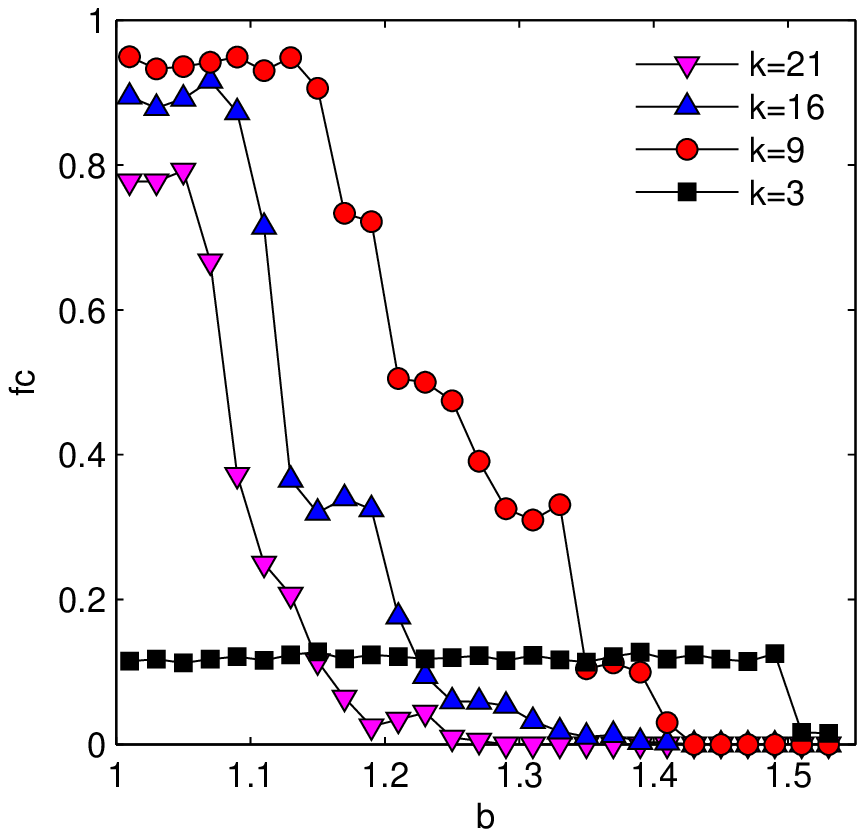}}
\subfigure[v=0.1]{\label{fc-b:subfig:b}
\includegraphics[height=5cm,width=5cm]{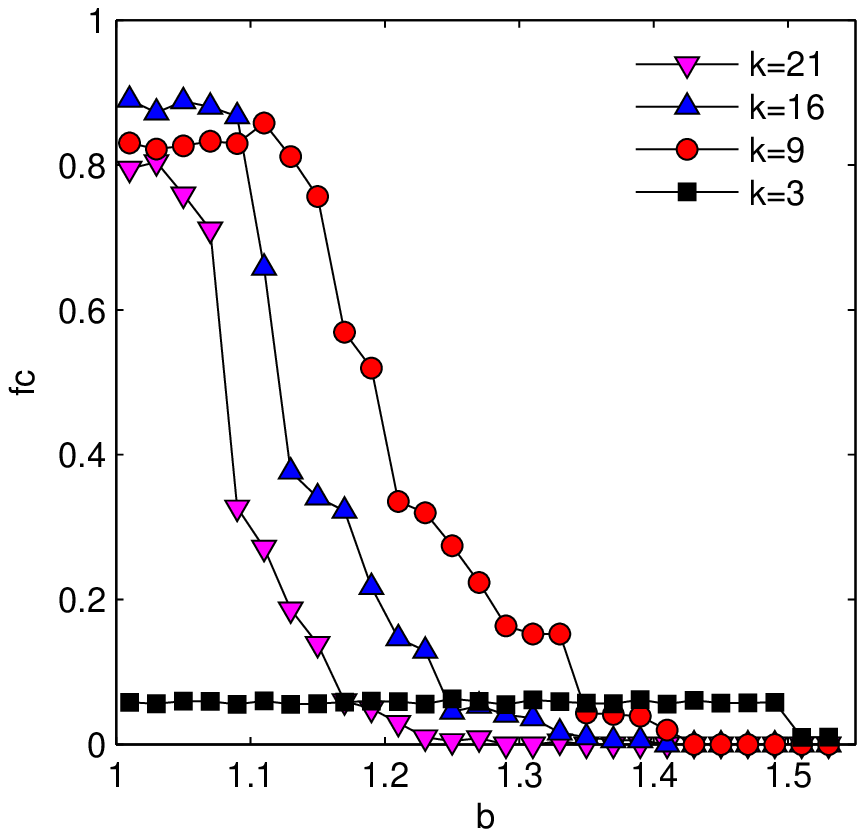}}
\subfigure[v=0.15]{\label{fc-b:subfig:c}
\includegraphics[height=5cm, width=5cm]{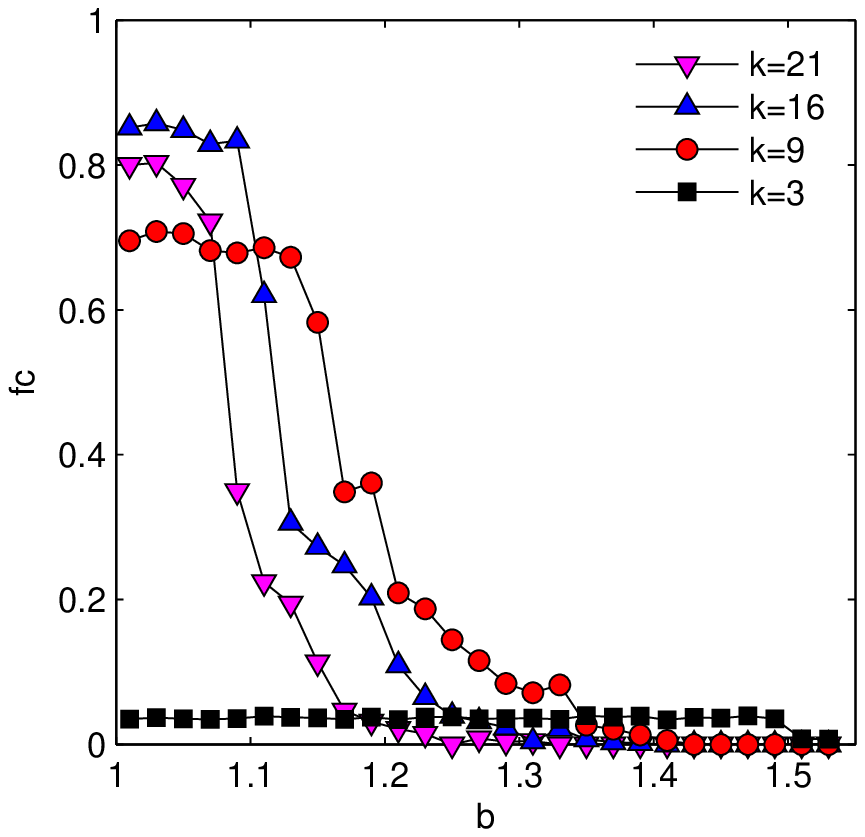}}
\subfigure[v=0.35]{\label{fc-b:subfig:d}
\includegraphics[height=5cm, width=5cm]{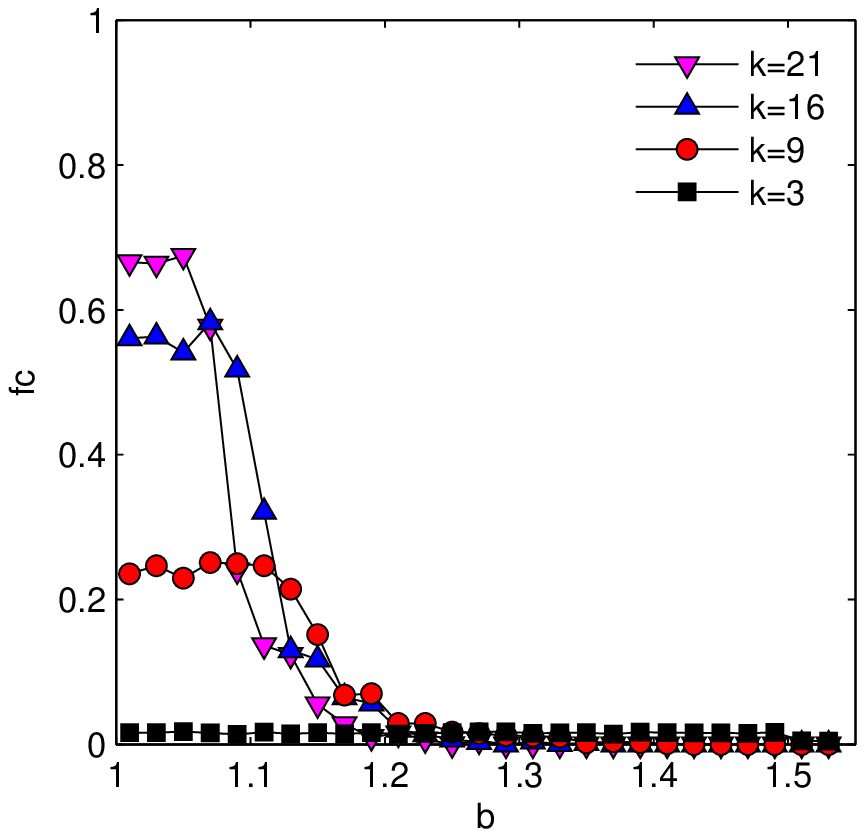}}
 \caption{The frequencies
of cooperators $fc$ versus the temptation to defect $b$ for
$v=0.05$, $0.1$, $0.15$ and $0.35$ respectively, where the cases
$k=21, 15, 9, 3$ correspond to different sizes of neighborhood, and
$b$ ranges from $1.01$ to $1.53$ with an interval of $0.02$. }
\label{fc-b:subfig}
\end{figure}

Fig. \ref{fc-b:subfig} illustrates the dependence of the frequency
of cooperators $fc$ on the temptation $b$ in the stationary state
for different sizes $k$ of neighborhood with a fixed absolute
velocity $v$. Under a fixed $v$ and $k$, $fc$ shows a step
structure, and gradually decreases as the increase of $b$. However,
the size $k$ of neighborhood can strongly affect the evolution of
cooperation. One can see that cooperators are more prone to die out
in the case of large $k$. But before the system is completely
occupied by defectors, there exists an appropriate $k$ to promote
cooperation for a fixed $b$. As shown in Fig. \ref{fc-b:subfig:a},
the cooperation level for $k=9$ is always higher than that for other
values of $k$, if $b<1.35$. If $1.35<b<1.51$, the highest level of
cooperation can be achieved when $k=3$. If $b>1.51$, defectors
dominate the population, no matter the values of $k$. These findings
suggest a non-monotonous dependence of the cooperator frequency on
the neighborhood size $k$. Besides, the absolute velocity $v$ also
plays an important role in the evolution of cooperation. Comparing
Fig. \ref{fc-b:subfig:b} with Fig. \ref{fc-b:subfig:a}, one can find
that the increase of $v$ leads an apparent drop of $fc$ for $k=9$ or
$k=3$. But for $k=15$ or $k=21$, the increment of $v$ does not cause
many changes to the cooperation level. And when $v=0.35$, the
highest level of cooperation for $k=21$ is still above $0.6$. These
findings imply that as the variance of $k$, the movement of
individuals has different influence on $fc$.

\begin{figure}[htbp]
\centering \subfigure[b=1.05]{ \label{fc-k:subfig:a}
\includegraphics[height=5cm,width=5cm]{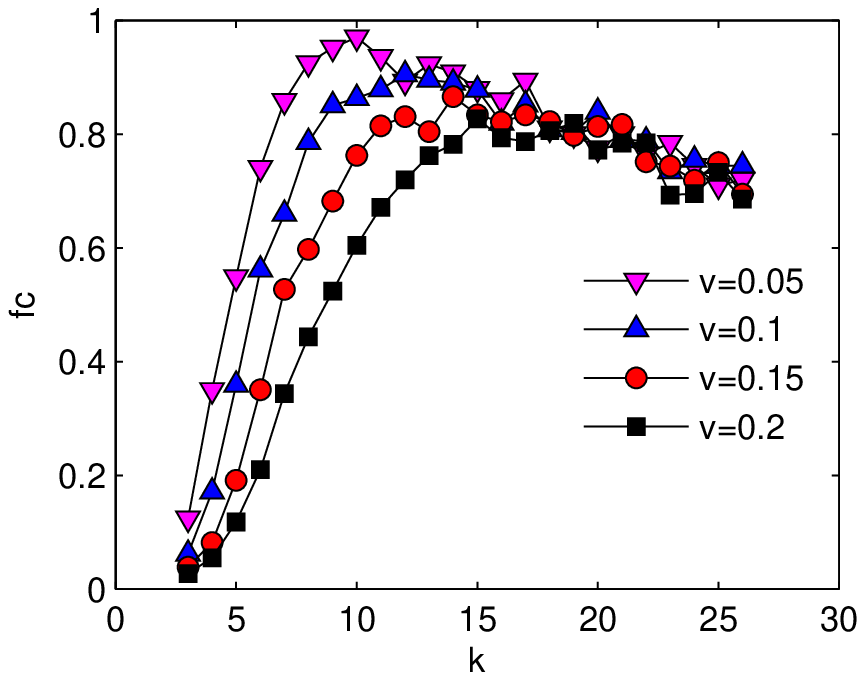}}
\subfigure[b=1.2]{\label{fc-k:subfig:b}
\includegraphics[height=5cm,width=5cm]{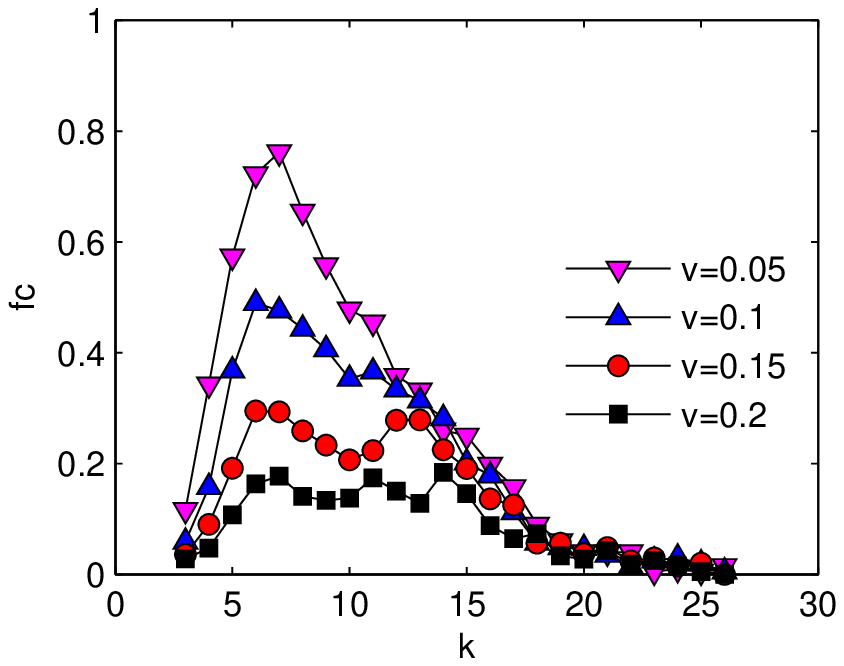}}
\caption{The frequency of cooperators $fc$ as a function of the size
of neighborhood $k$ for $b=1.05$ and $b=1.2$
respectively. And $k$ ranges from $3$ to $26$ with an interval of $1$.}\label{fc-k:subfig}
\end{figure}

To investigate the role of the neighborhood size $k$, Fig.
\ref{fc-k:subfig} presents the cooperator frequency $fc$ as a
function of the neighborhood size $k$ for a fixed temptation $b$.
Dai et al. reported the promotion of cooperation through enlarging
the size of neighborhood among mobile agents \cite{Dai2007}, where
molecular dynamics is used to describe repulsion and attraction
between agents in flocks. But in our model, a resonance-like
behavior can be observed: there exists a peak of $fc$ at some values
of $k$. In fact, the same behavior has been found in three typical
networks, where the density of cooperators peaks at some specific
values of the average degree \cite{Tang2006}. Here our work can be
viewed as extensions of previous work to dynamical networks. Next we
will give a simple explanation for the non-trivial relation between
$fc$ and $k$. On square lattices and regular ring-graphs, fixed
locations of players provide continuous interactions within local
neighborhoods, and cooperators can cluster together to resist the
invasion of defectors \cite{Szabo1998,Santos2005}. The increment of
average degree indeed hampers cooperation, because the well-mixed
limit is nicely approached for a sufficiently large size of
neighborhood \cite{Santos2006a}. When players are kept moving,
however, the cluster of cooperators may be destroyed by time-variant
neighborhoods. The smaller $k$ is, the longer time the system needs
to form a fixed interaction network. The increment of $k$ enhances
the probability of future encounters between players and their
former neighbors. As a result, interactions among cooperators can be
maintained. But defectors can also exploit more cooperators as $k$
increases, and large values of $k$ reproduce the mean field
situation. To promote cooperation, there should be a compromise
between the two limits of $k$ discussed above. That is why the
cooperation level reaches the maximum only at intermediate values of
$k$. At the same time, the positive effect coming from intermediate
local connections on cooperation is greatly constrained by the
absolute velocity $v$ and the temptation $b$. In Fig.
\ref{fc-k:subfig},  the value of $fc$ at the peak point decreases as
$v$ increases. And when $b$ increases to $1.2$ in Fig.
\ref{fc-k:subfig:b}, the curve of $fc$ is almost leveled off for
$v=0.2$.

\begin{figure}[htbp]
\centering \subfigure[]{ \label{fcv:n09}
\includegraphics[height=5cm,width=5cm]{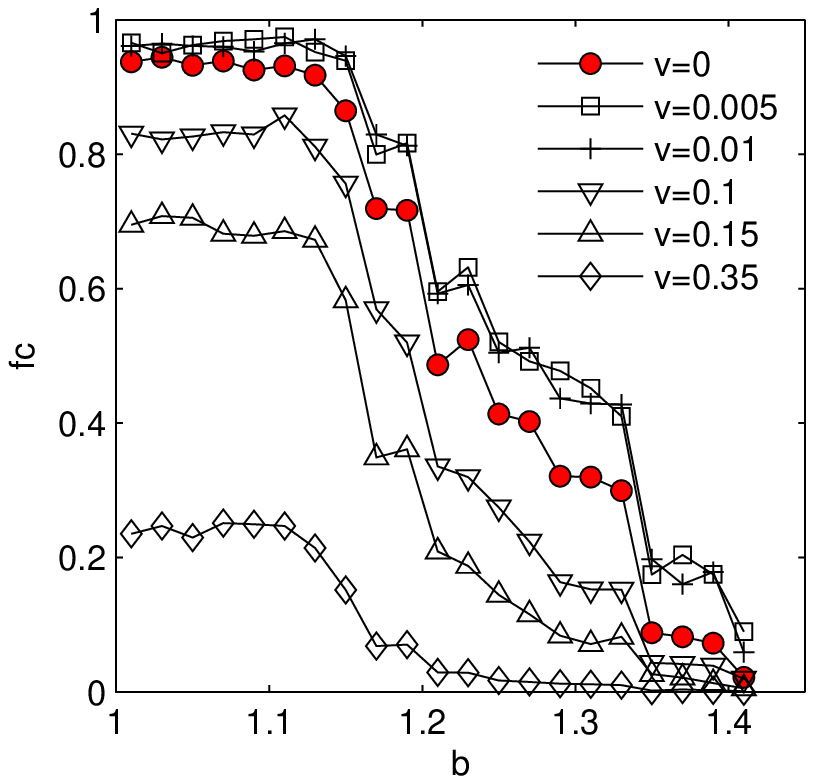}}
\subfigure[]{\label{fcv:b117}
\includegraphics[height=5cm,width=5cm]{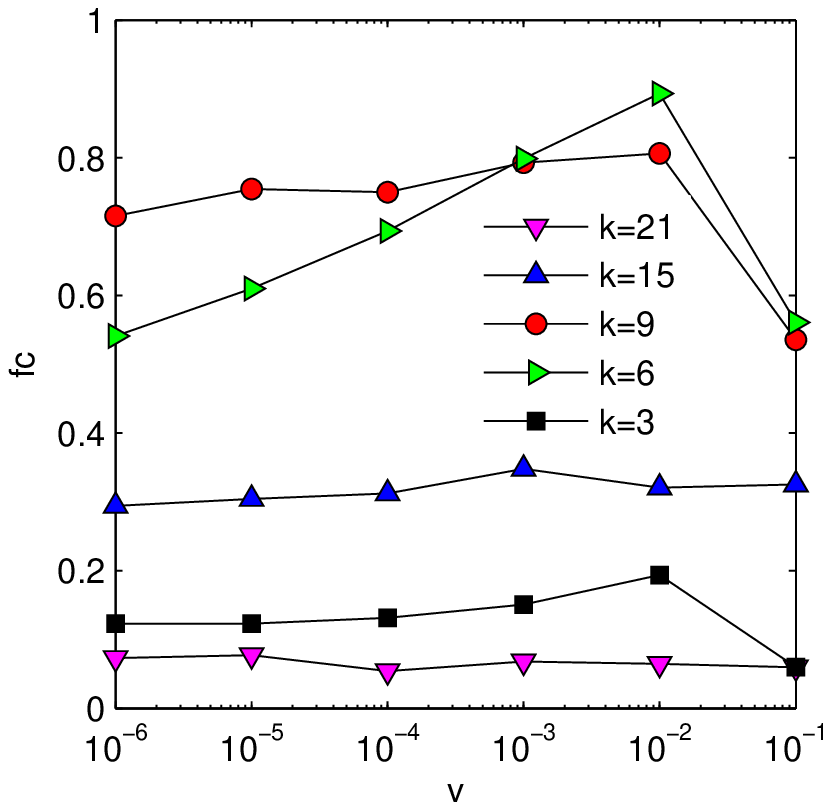}}
\caption{(a) The frequency of cooperators $fc$ as a function of the
temptation $b$ for $k=9$ with various velocities $v$. $b$ ranges
from $1.01$ to $1.41$ with an interval of $0.02$. (b) The frequency
of cooperators $fc$ versus the absolute velocity $v$ for $b=1.17$
with different sizes $k$ of neighborhood. A logarithmic scale is
used for the $X$ axis.}\label{fcv}
\end{figure}

To study the effect of the absolute velocity $v$ on the cooperation
level, Fig. \ref{fcv:n09} demonstrates the frequency of cooperators
$fc$ as a function of the temptation $b$ for different values of $v$
with $k=9$. In our model, velocity $v$ measures the movement speed
of players. One can find that for $v\geq 0.1$, the cooperator
frequency is lower than that for $v=0$, and decreases gradually as
$v$ increases. In fact, when the agents move with a high velocity,
they have greater chance to contact with different neighbors than
that in the case of small $v$. Before the interaction network gets
fixed, neighbors of each agent change quite often, or might be
completely different at each time step. As a result, there is a
small probability of forming compact clusters of cooperators, which
leads to the dominance of defectors. For $v\leq 0.01$, however, the
situation is reversed. Compared with the case that agents do not
move, the cooperation level is promoted throughout the whole
parameter range of $b$ when $v=0.005$ or $v=0.01$. It suggests that
cooperation among mobile individuals is not only possible but may
even be enhanced, and this finding is in accordance with the
previous work \cite{Vainstein2007,Dai2007,Meloni2009}. But such
effect relies on the size $k$ of neighborhood, as shown in Fig.
\ref{fcv:b117}, which presents the dependence of $fc$ on $v$ for
different values of $k$ with $b=1.17$. For $k\leq 9$, the
cooperation level increases with $v$, and reaches the maximum value
around $v=0.01$. When $v=0.1$, a drop of $fc$ appears. For $k=15$ or
$k=21$, $fc$ changes little when $v$ increases. This can be
explained by the occurrence of mean field situation at large values
of $k$, which offsets the enhancement of cooperation from mobility.

\begin{figure}[htbp]
\centering
\subfigure[]{\label{dens-v}
\includegraphics[height=5cm, width=5cm]{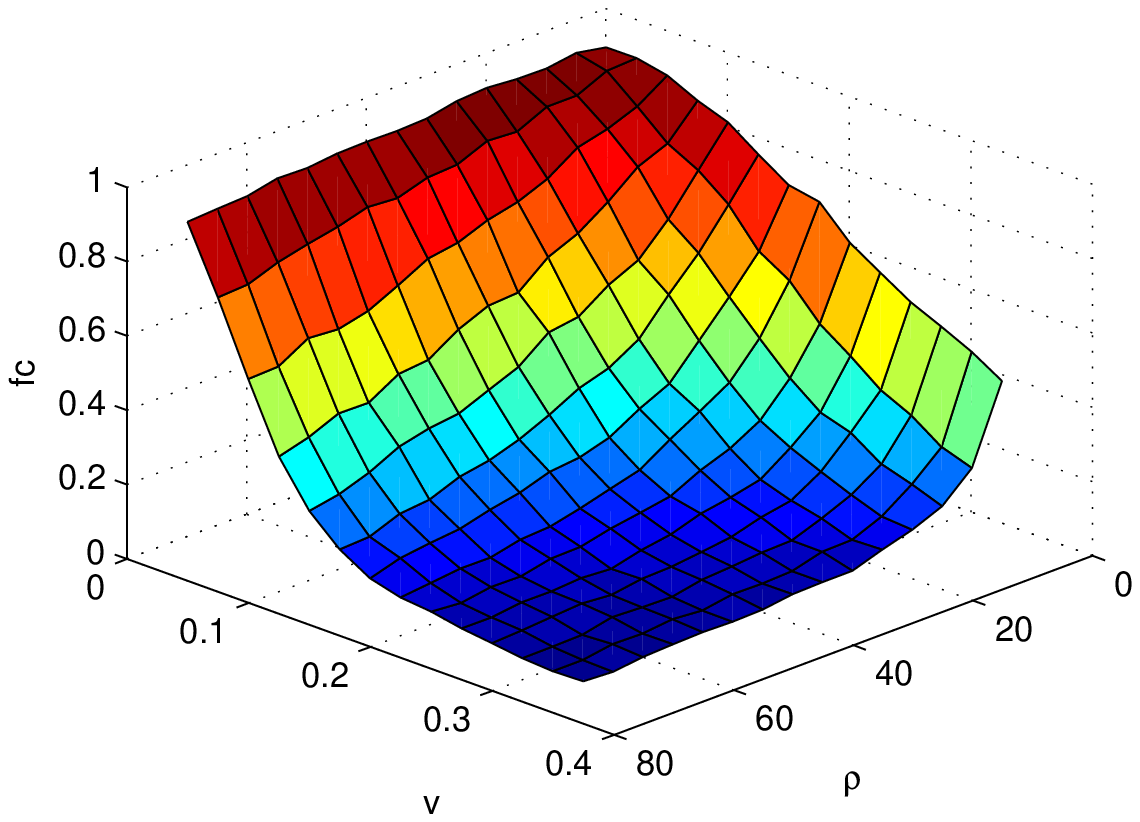}}
\subfigure[]{\label{dens-k}
\includegraphics[height=5cm, width=5cm]{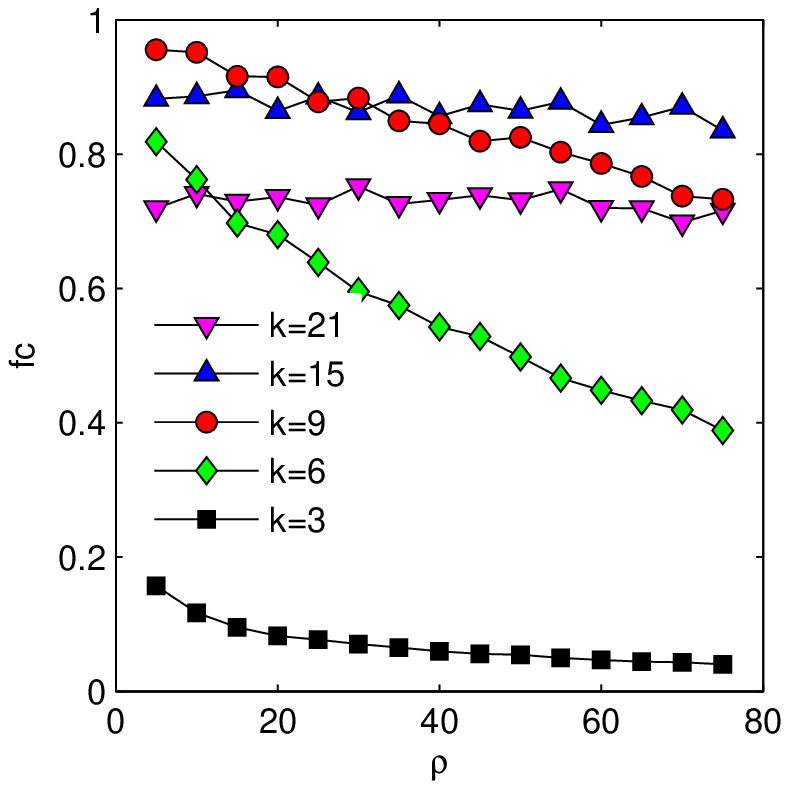}}
\caption{(a) The cooperator frequencies $fc$ versus the absolute
velocity $v$ and the density $\rho$ for $k=9$, $b=1.07$. (b) The
cooperator frequencies $fc$ versus the density $\rho$ for $v=0.05$,
$b=1.05$. $v$ ranges from $0.025$ to $0.35$ with an interval of
$0.025$, and $\rho$ ranges from $5$ to $75$ with an interval of
$5$.} \label{dens}
\end{figure}

The evolution of system relies on the density $\rho$ of agents at
$t=0$, which can be defined as $\rho=N/L^2$. And it has been
reported that there is a optimal region of $\rho$ for cooperation,
when the neighbors are chosen according to a prescribed distance
\cite{Meloni2009}. Fig. \ref{dens-v} shows the combining effect of
$v$ and $\rho$ on the cooperator frequency $fc$ for $k=9$ and
$b=1.07$. For a fixed $v$, one can find that $fc$ decreases
monotonously as $\rho$ increases, and the decreasing velocity
increases with $v$. Clearly, our finding is different with that
reported in Ref. \cite{Meloni2009}, and this difference is rooted in
the definition of neighborhoods. Here $\rho$ indicates how dense
players distribute on the plane when $t=0$. In Ref.
\cite{Meloni2009}, $\rho$ determines the average degree $<k>$ of the
interaction network, and $<k>$ increases with $\rho$. Previous work
has revealed that moderate values of average degree can enhance
cooperation \cite{Tang2006}. Then the existence of the optimal
region of $\rho$ for cooperation becomes understandable. In our
model, however, each agent plays with a constant number of
neighbors. The increasing of $\rho$ produces a dense population,
which brings fast change in neighborhoods for the players. And for a
fixed $\rho$, a sufficiently large $v$ would hamper the evolution of
cooperation, as discussed above. Hence the system shows low values
of $fc$ for large $\rho$ and $v$. Fig. \ref{dens-k} sheds more light
on the role of $k$ when $\rho$ increases. When $k\leq 9$, the
increase of $\rho$ leads an apparent decrease of $fc$. While for
$k\geq 21$, variation of densities only causes small fluctuations of
$fc$.

\begin{figure}[htbp]
\centering \subfigure[t=0]{ \label{agg:subfig:a}
\includegraphics[height=4cm,width=3.5cm]{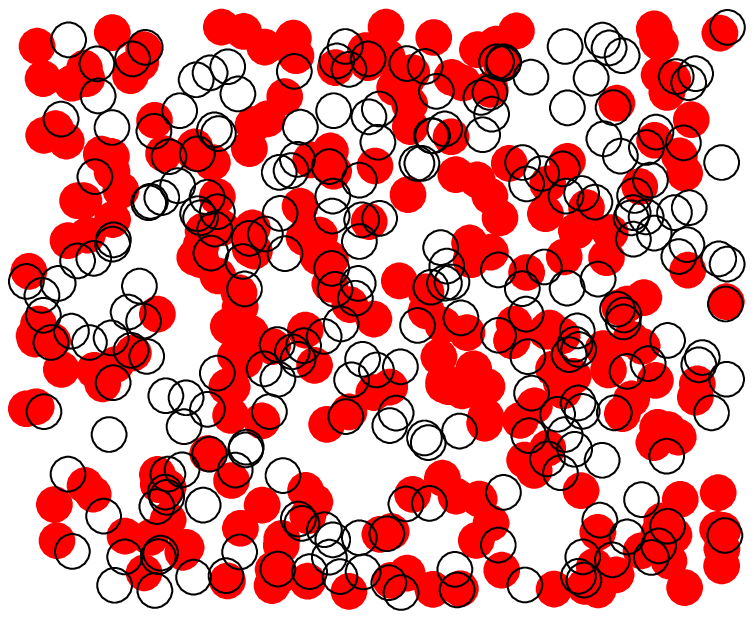}}
\subfigure[t=25]{\label{agg:subfig:b}
\includegraphics[height=4cm,width=3.5cm]{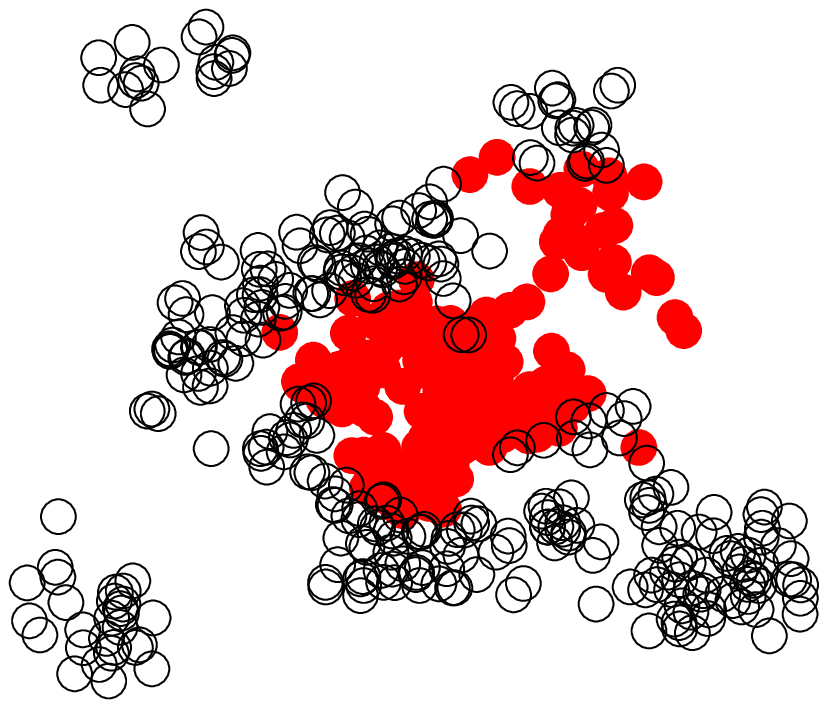}}
\subfigure[t=100]{\label{agg:subfig:c}
\includegraphics[height=4cm, width=3.5cm]{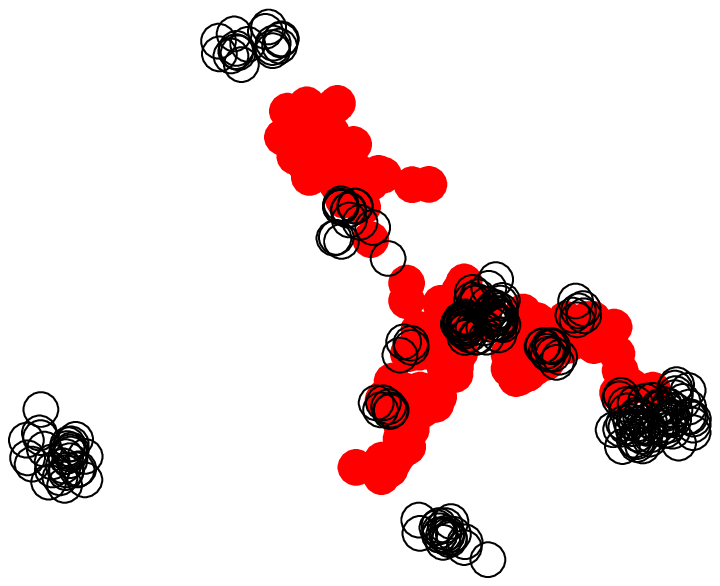}}
\subfigure[t=1300(equilibrium)]{\label{agg:subfig:d}
\includegraphics[height=5cm, width=5cm]{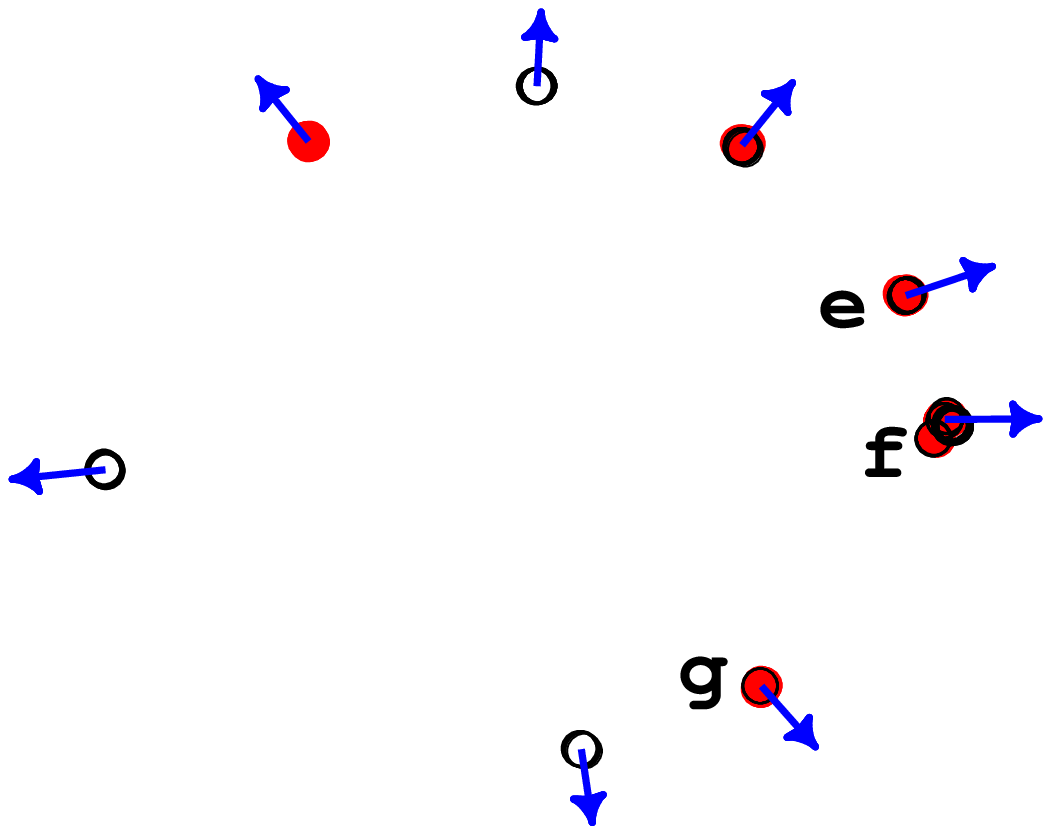}}
\subfigure[]{\label{agg:subfig:e}
\includegraphics[height=5cm, width=5cm]{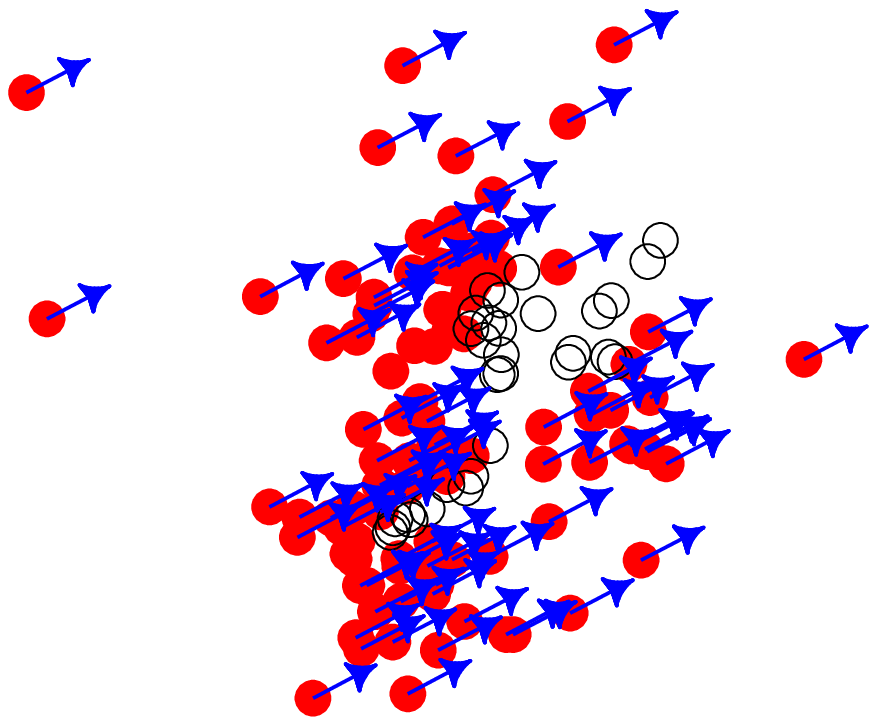}}
\subfigure[]{\label{agg:subfig:f}
\includegraphics[height=5cm, width=5cm]{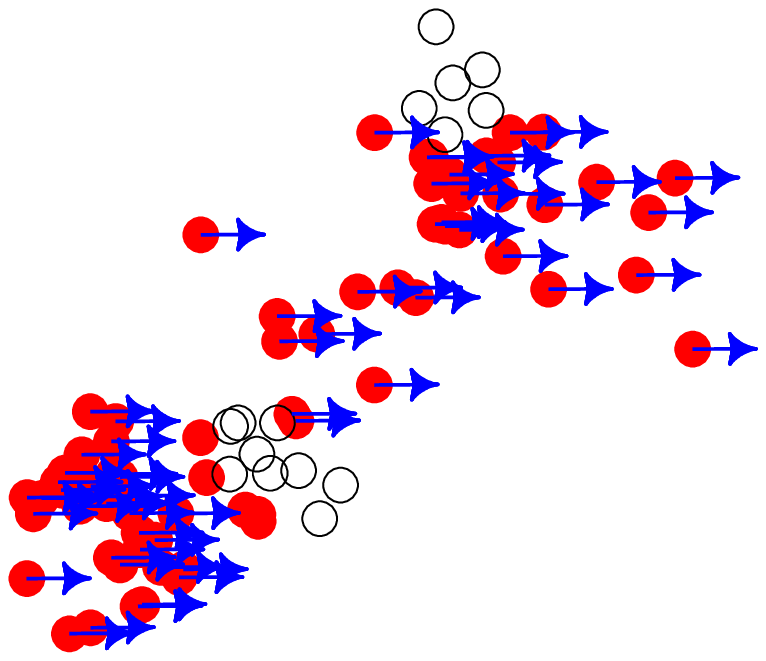}}
\subfigure[]{\label{agg:subfig:g}
\includegraphics[height=5cm, width=5cm]{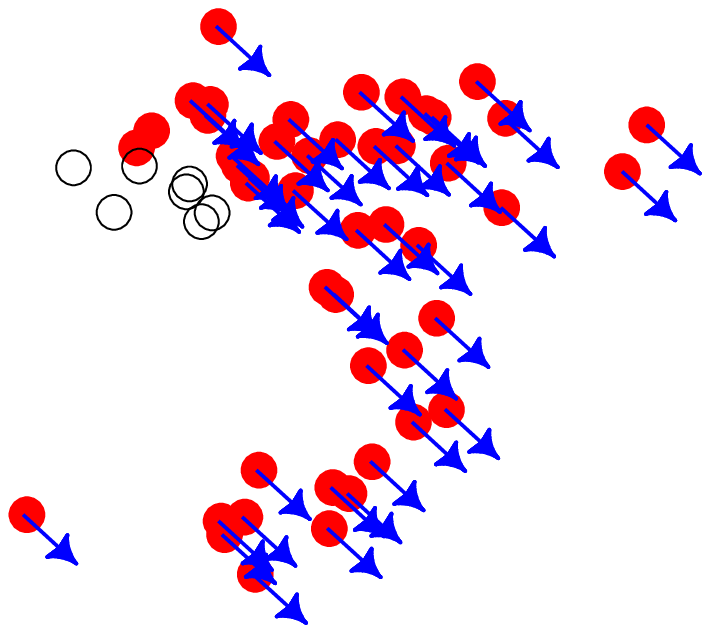}}
 \caption{Snapshots of the evolution of cooperation with $b=1.35$, $k=9$ and $v=0.05$.
 Cooperators (red circles) form clusters to resist the invasion of defectors (white circles).
 At an equilibrium state, players running toward the same direction stay together,
 and their velocities are denoted by arrows.
 The last three figures present details of the labeled components in (d).
 To give a clear figure of spatial configuration,
 not all directions of the agents are denoted in (e), (f) and (g).} \label{agg:subfig}
\end{figure}

To have an insight into the evolution of cooperation among mobile
players, Fig. \ref{agg:subfig} provides snapshots of spatial
configurations at equilibrium, which is obtained in one realization.
And to eliminate additional mechanisms that favor cooperation, the
value of the temptation is near the extinction threshold of
cooperators. One can find that the system gradually splits into many
small flocks, in which all the agents move toward a same direction.
Because the agents are located  in a plane without boundary
restrictions, they fly apart and never meet again. During the
process of direction alignment, cooperators can survive by forming
compact clusters. And the two strategies may coexist at equilibrium,
as shown in the last three figures. One can see that defectors are
located on the border of flocks, or surrounded by cooperators. For
cooperators adjacent to defectors, mutual cooperation make their
cooperative neighbors earn higher payoffs than the income of
defectors. According to the best-takes-over rule of strategy update,
the cooperators will follow the strategies of their cooperative
neighbors. That is why cooperation can be maintained in population,
and this mechanism has been found in the lattice structure
\cite{Nowak1993}.

\section{Conclusion}
To summarize,
we investigate the effects of mobility on the evolution of cooperation in the direction alignment process of flocks.
Numerical simulations show that cooperation can be maintained in mobile players with simple strategies.
Depending on the temptation to defect and the velocity at which the agents move,
there exist an optimal size of interaction neighborhood to produce the maximum cooperation level.
When compared with the case that all agents do not move,
the cooperation level can even be enhanced by the mobility of individuals,
if the velocity and the size of neighborhood are small.
The cooperation level is also affected by the density $\rho$ of agents, and $fc$ decreases as the increase of $\rho$.
Moreover, the system exhibits aggregation behavior, and we illustrate the coexistence of different strategies at equilibrium.
Our work may be relevant for understanding the role of information flows in cooperative, multi-vehicle systems \cite{Olfati-Saber2007}.

This work is supported by the Key Fundamental Research Program of Shanghai (Grant No.09JC1408000), the National Key Fundamental Research
Program (Grant No.2002cb312200) and the National Natural Science
Foundation of China (Grant No.60575036).


\begin{thebibliography}{00}
\bibitem{Hauert2004}
C.~Hauert, M.~Doebeli, Spatial structure often inhibits the evolution of
  cooperation in the snowdrift game, Nature 428 (2004) 643--646.

\bibitem{Nowak1993}
M.~A. Nowak, R.~M. May, The spatial dilemmas of evolution, Int. J. Bifurcation
  Chaos 3 (1993) 35--78.

\bibitem{Szabo1998}
G.~Szab{\'o}, C.~T{\H o}ke, Evolutionary prisoner's dilemma game on a square
  lattice, Phys. Rev. E 58 (1998) 69--73.

\bibitem{Perc2008}
M.~Perc, A.~Szolnoki, Social diversity and promotion of cooperation in the
  spatial prisoner's dilemma game, Phys. Rev. E 77 (2008) 011904.

\bibitem{Wu2006}
Z.~X. Wu, X.~J. Xu, Z.~G. Huang, S.~J. Wang, Y.~H. Wang, Evolutionary
  prisoner's dilemma game with dynamic preferential selection, Phys. Rev. E 74
  (2006) 021107.

\bibitem{Santos2005}
F.~C. Santos, J.~M. Pacheco, Scale-free networks provide a unifying framework
  for the emergence of cooperation, Phys. Rev. Lett. 95 (2005) 098104.

\bibitem{Gomez-Gardenes2007}
J.~Gomez-Gardenes, M.~Campillo, L.~M. Floria, Y.~Moreno, Dynamical organization
  of cooperation in complex topologies, Phys. Rev. Lett. 98 (2007) 108103.

\bibitem{Ren2007}
J.~Ren, W.~X. Wang, F.~Qi, Randomness enhances cooperation: A resonance-type
  phenomenon in evolutionary games, Phys. Rev. E 75 (2007) 045101.

\bibitem{Chen2008}
X.~J. Chen, L.~Wang, Promotion of cooperation induced by appropriate payoff
  aspirations in a small-world networked game, Phys. Rev. E 77 (2008) 017103.

\bibitem{Du2009}
W.~B. Du, X.~B. Cao, M.~B. Hu, H.~X. Yang, H.~Zhou, Effects of expectation and
  noise on evolutionary games, Physica A 388 (2009) 2215--2220.

\bibitem{Szabo2007}
G.~Szab{\'o}, G.~F{\'a}th, Evolutionary games on graphs, Phys. Rep. 446 (2007)
  97--216.

\bibitem{Nowak2006}
M.~A. Nowak, Five rules for the evolution of cooperation, Science 314~(5805)
  (2006) 1560--1563.

\bibitem{Doebeli2005}
M.~Doebeli, C.~Hauert, Models of cooperation based on the prisoner's dilemma
  and the snowdrift game, Ecol. Lett. 8~(7) (2005) 748--766.

\bibitem{Zimmermann2004}
M.~G. Zimmermann, V.~M. Egu{\'\i}luz, M.~San~Miguel, Coevolution of dynamical
  states and interactions in dynamic networks, Phys. Rev. E 69 (2004) 065102.

\bibitem{Santos2006}
F.~C. Santos, J.~M. Pacheco, T.~Lenaerts, Cooperation prevails when individuals
  adjust their social ties, PLOS Comp. Biol. 2 (2006) 1284--1291.

\bibitem{Szolnoki2008a}
A.~Szolnoki, M.~Perc, Z.~Danku, Making new connections towards cooperation in
  the prisoner's dilemma game, EPL 84~(5) (2008) 50007.

\bibitem{Szolnoki2008}
A.~Szolnoki, M.~Perc, Coevolution of teaching activity promotes cooperation,
  New J. Phys. 10 (2008) 043036.

\bibitem{Szolnoki2009}
A.~Szolnoki, M.~Perc, Promoting cooperation in social dilemmas via simple
  coevolutionary rules, Eur. Phys. J. B 67~(3) (2009) 337--344.

\bibitem{Fu2008}
F.~Fu, C.~Hauert, M.~A. Nowak, L.~Wang, Reputation-based partner choice
  promotes cooperation in social networks, Phys. Rev. E 78 (2008) 026117.

\bibitem{Gonzalez2006}
M.~C. Gonz{\'a}lez, P.~G. Lind, H.~J. Herrmann, System of mobile agents to
  model social networks, Phys. Rev. Lett. 96 (2006) 088702.

\bibitem{Brockmann2006}
D.~Brockmann, L.~Hufnagel, T.~Geisel, The scaling laws of human travel, Nature
  439 (2006) 462--465.

\bibitem{Gonzalez2008}
M.~C. Gonz{\'a}lez, C.~A. Hidalgo, A.~L. Barab{\'a}si, Understanding individual
  human mobility patterns, Nature 453 (2008) 779--782.

\bibitem{LeGalliard2005}
J.~F. Le~Galliard, R.~Ferri{\`e}re, U.~Dieckmann, Adaptive evolution of social
  traits: Origin, trajectories, and correlations of altruism and mobility, Am.
  Nat. 165 (2005) 206--224.

\bibitem{Hamilton2005}
I.~M. Hamilton, M.~Taborsky, Contingent movement and cooperation evolve under
  generalized reciprocity, Proc. R. Soc. B 272 (2005) 2259--2267.

\bibitem{Aktipis2004}
C.~A. Aktipis, Know when to walk away: contingent movement and the evolution of
  cooperation, J. Theor. Biol. 231 (2004) 249--260.

\bibitem{Helbing2009}
D.~Helbing, W.~J. Yu, The outbreak of cooperation among success-driven
  individuals under noisy conditions, Proc. Natl. Acad. Sci. USA 106~(10)
  (2009) 3680--3685.

\bibitem{Vainstein2007}
M.~H. Vainstein, A.~T.~C. Silva, J.~J. Arenzon, Does mobility decrease
  cooperation?, J. Theor. Biol. 244 (2007) 722--728.

\bibitem{Dai2007}
X.~B. Dai, Z.~Y. Huang, C.~X. Wu, Evolution of cooperation among interacting
  individuals through molecular dynamics simulations, Physica A 383 (2007)
  624--630.

\bibitem{Meloni2009}
S.~Meloni, A.~Buscarino, L.~Fortuna, M.~Frasca, J.~Gomez-Gardenes, V.~Latora,
  Y.~Moreno, Effects of mobility in a population of prisoner's dilemma players,
  Phys. Rev. E 79 (2009) 067101.

\bibitem{Dossetti2009}
V.~Dossetti, F.~J. Sevilla, V.~M. Kenkre, Phase transitions induced by complex
  nonlinear noise in a system of self-propelled agents, Phys. Rev. E 79 (2009)
  051115.

\bibitem{Vicsek1995}
T.~Vicsek, A.~Czir{\'o}k, E.~Ben-Jacob, I.~Cohen, O.~Shochet, Novel type of
  phase transition in a system of self-driven particles, Phys. Rev. Lett. 75
  (1995) 1226--1229.

\bibitem{Tang2006}
C.~L. Tang, W.~X. Wang, X.~Wu, B.~H. Wang, Effects of average degree on
  cooperation in networked evolutionary game, Eur. Phys. J. B 53~(3) (2006)
  411--415.

\bibitem{Santos2006a}
F.~C. Santos, J.~F. Rodrigues, J.~M. Pacheco, Graph topology plays a
  determinant role in the evolution of cooperation, Proc. R. Soc. B 273 (2006)
  51--55.

\bibitem{Olfati-Saber2007}
R.~Olfati-Saber, J.~A. Fax, R.~M. Murray, Consensus and cooperation in
  networked multi-agent systems, Proc. IEEE 95 (2007) 215--233.

\end{thebibliography}
\end{document}